\documentclass[twocolumn,english,pre]{revtex4}
\usepackage[T1]{fontenc}
\usepackage[latin1]{inputenc}
\usepackage{amsmath}
\usepackage{graphicx}
\usepackage{amssymb}

\makeatletter

\providecommand{\LyX}{L\kern-.1667em\lower.25em\hbox{Y}\kern-.125emX\@}

\usepackage{amsfonts}

\providecommand{\LyX}{L\kern-.1667em\lower.25em\hbox{Y}\kern-.125emX\@}

\usepackage{babel}
\makeatother
\begin{document}

\title{Variance of transmitted power in multichannel dissipative ergodic
structures invariant under time reversal}

\author{Igor Rozhkov$^{1}$, Yan V. Fyodorov$^{2,3}$, Richard L. Weaver$^{1}$}

\begin{abstract}
We use random matrix theory (RMT) to study the first two moments of
the wave power transmitted in time reversal invariant systems having
ergodic motion. Dissipation is modeled by a number of loss channels
of variable coupling strength. To make a connection with ultrasonic
experiments on ergodic elastodynamic billiards, the channels injecting
and collecting the waves are assumed to be negligibly coupled to the
medium, and to contribute essentially no dissipation. Within the RMT
model we calculate the quantities of interest exactly, employing the
supersymmetry technique. This approach is found to be more accurate
than another method based on simplifying naive assumptions for the
statistics of the eigenfrequencies and the eigenfunctions. The results
of the supersymmetric method are confirmed by Monte Carlo numerical
simulation and are used to reveal a possible source of the disagreement
between the predictions of the naive theory and ultrasonic measurements. 
\end{abstract}

\affiliation{$^{1}$Department of Theoretical and Applied Mechanics, University
of Illinois at Urbana-Champaign, Urbana, IL 61801\\
 $^{2}$Department of Mathematical Sciences, Brunel University, Uxbridge
UB83PH, United Kingdom\\
 $^{3}$Petersburg Nuclear Physics Institute RAN, Gatchina 188350,
Russia}

\maketitle

\section{Introduction}

The statistics of waves in complex disordered and ray-chaotic structures
have been well modeled in recent years by random matrix theory (RMT).
The eigenstatistics of such structures are ergodically equivalent
to those of certain classes of random matrices. This has been established
by an enormous literature, both experimental and theoretical, and
applies to the complex dynamics of compound nuclei \cite{Porter},
and also to the somewhat simpler case of closed nondissipative wave
billiards with chaotic ray trajectories \cite{BGS}. More recently
attention has turned to the case of structures with open loss channels
and/or internal dissipation {[}3--8{]}. This evolution of focus has
been dictated by the physical systems available, for which it is difficult
or impossible to eliminate absorption, and/or minimize the dissipative
effect of the channels used to launch or detect the waves.

Of the many statistics that could be considered for such structures,
perhaps the simplest experimentally accessible one is the relative
variance of the power transmission. This quantity is related to cross
section fluctuations in nuclear reactions, it is accessible in microwave
experiments, and is of long standing interest in acoustics \cite{Lyon,LobkisJSV}.
In Ref. \cite{LobkisJSV} relative variances measured in a dissipative
3D ultrasonic billiard were compared to the predictions of a simple
theory which assumes, that the eigenstatistics are identical to those
of the nondissipative case. Such an assumption is strictly true only
for very special cases of dissipation. The theory was found to consistently
overestimate the relative variance of the mean square transmitted
signal. Our chief interest here is to develop a more rigorous theory
for that variance, and to compare it with the predictions of the naive
theory. 

As an adequate theoretical model of such a structure we will use an
effective random matrix theory description, with a random matrix $H$
replacing the wave equation's linear differential operator (see \cite{GUEpaper}
and references therein for a more detailed discussion). Then the complex
amplitude of the transmitted wave is simply proportional to the $ij$th
matrix element of the resolvent associated with the wave equation:
$G(E)\equiv \left[E\, I+i\varepsilon \, I-H+i\Gamma \right]^{-1}$.
Here, $I$ is the identity matrix, the matrix $\Gamma $ describes
coupling to internal local-in-space dissipative channels, the parameter
$\varepsilon >0$ describes uniform losses, and $E$ is the spectral
variable. The same model describes microwave billiards, ultrasonic
billiards and nuclear reactions. The real symmetric positive semi-definite
loss operator $\Gamma $ can be written in terms of the states of
the channels ($\Gamma =W^{\dagger }W$ in definitions of Ref. \cite{VWZ})
or in terms of absorption mechanisms. It is generally taken to be
only weakly dependent on $E$. Thus both open and closed dissipative
systems are described by the same model. When losses are negligible
the experimental systems are usually invariant under time reversal.
The appropriate choice for the corresponding random matrix $H$ should,
therefore, be a real symmetric matrix taken from the Gaussian Orthogonal
Ensemble (GOE). 

Our quantity of prime interest is $T=\left|G_{ij}\left(E\right)\right|^{2},i\neq j$,
i.e. the product of retarded and advanced Greens functions (propagators):
$G_{ji}^{R}(E)\equiv \left[E\, I+i\varepsilon \, I-H+i\Gamma \right]_{ji}^{-1}$
and $G_{ij}^{A}=\left(G_{ji}^{R}(E)\right)^{\ast }$ respectively.
Except for slowly varying factors of receiver gain and source strength,
the quantity $T$ represents the ultrasonic power of Ref. \cite{LobkisJSV}. 

For general non-perturbative statistical studies the only generally
applicable tool known at present is reduction to the Efetov's zero
dimensional supersymmetric non-linear $\sigma $-model \cite{Efetovbook}.
In this way the problem of calculating RMT ensemble averages reduces
to performing a definite finite-dimensional integral over a space
of supermatrices. The zero dimensional $\sigma $-model can also be
derived from the assumptions of RMT \cite{VWZ}, and is applicable
to a variety of quantum-scattering problems formulated in terms of
random Hamiltonians {[}8-12{]}. Ideally, once the quantity of interest
is expressed in terms of the products of resolvents of the effective
Hamiltonian $H-i\Gamma $, its mean, its variance, and sometimes its
distribution function can be obtained. 

Technical details of the supersymmetric reduction procedure depend
essentially on the basic symmetries of the underlying ensemble. It
is well known, that working with the orthogonal ensemble leads to
calculations, which are more technically involved than those of similar
calculations for systems with broken time-reversal invariance (TRI).
In the latter case, the corresponding ensemble is the Gaussian Unitary
(GUE), and one can go as far as calculating the full distribution
function of transmitted wave power in ergodic systems \cite{GUEpaper}.
Unfortunately, the existing experimental results on power transmission
are only available for systems with preserved TRI. 

The main goal of the present work is to explore transmitted power
statistics for ergodic TRI systems. We find that it is possible to
derive explicit analytical expressions for the first two moments of
this quantity. We wish in particular to explain the differences, seen
in \cite{LobkisJSV}, between the predictions of the oversimplified
({}``naive'' perturbative) theory and experimental measurements. 

In section II we use the supersymmetry method to derive expressions
for mean and variance of transmitted power. In Section III, we confirm
the results by numerical Monte Carlo calculation, and compare them
with the results of the perturbative method of Refs. \cite{GUEpaper,LobkisJSV}.
In Section IV we investigate a hypothesis to explain the longstanding
discrepancy between lab measurements in ergodic acoustic systems and
naive RMT predictions. A summary is given in Section V.

\section{Supersymmetric calculation for mean and mean squared power}

\subsection{The system}

In an ergodic system characterized by a random $N\times N$ Hamiltonian
$H$ and a dissipation matrix $\Gamma $, a matrix element of $G\left(E\right)=\left(EI+i\varepsilon I-H+i\Gamma \right)^{-1}$
represents the response spectrum (with $E$ being the spectral variable).
Its squared absolute value $G_{ij}\left(E\right)G_{ij}^{\ast }\left(E\right)$
denotes the spectral power density.

The elements of the random matrix $H$ are zero-centered Gaussian
variables, and because we deal with with power transmission inside
time-reversal invariant systems, the matrix $H$ is real and symmetric.
The relevant random matrix ensemble is, therefore, the Gaussian orthogonal
ensemble (GOE). Because of the inherent orthogonal invariance the
dissipation matrix may be chosen to be diagonal: $\Gamma =\textrm{diag}\{\gamma ,\gamma ,...\gamma ,0,...0\}$,
as we always can express our matrices in $\Gamma $'s natural basis.
The number $M<N$ of nonzero entries can be interpreted either as
a number of equivalent open channels in the scattering system or,
as a number of equivalent localized {}``dampers'' in a closed system
with losses \cite{LobkisJSV}. Note, that convergence generating parameter
$\varepsilon >0$ can be interpreted as the coupling to infinite number
of external channels or, as uniform dissipation. 

We are interested in the statistics of the wave power $T$ transmitted
from a source at point $j$ to a receiver at a different point $i$
inside the system: $T=G_{ij}\left(E\right)G_{ij}^{\ast }\left(E\right),$
$1\leq i,j\leq N\, j\ne i$ (no summation over $i$ and $j$) \cite{LobkisJSV}.

\subsection{Basic definitions and identities}

To obtain expressions for the first two moments of the transmitted
power $\overline{T}$, $\overline{T^{2}}$ (the bar indicates the
ensemble averaging) we adopt a procedure similar to that of Ref. \cite{VWZ}.
We start with constructing a generating function $Z$ for our quantities
of interest by introducing $4$-component supervectors $\Phi _{p}$:\[
\Phi _{p}^{T}=\left\{ \chi _{p}^{\ast T},\chi _{p}^{T},S_{p}(1)^{T},S_{p}(2)^{T}\right\} ,\quad p=1,2\, ,\]
where the components of $N$-dimensional vectors $S$ are real commuting
variables, the elements of the vectors $\chi $ are anticommuting
variables (Grassmannian), and $^{T}$ stands for the vector transposition.
Index $p$ is used to distinguish between retarded ($p=1$) and advanced
($p=2$) Green's functions. The latter can be obtained from the generating
functions: $Z_{p}\left(E,\mathfrak{J}_{p}\right)=\int \left[d\Phi _{p}\right]\exp \left\{ \left(i/2\right)\mathfrak{L}_{p}\left(E,\Phi _{p},\mathfrak{J}_{p}\right)\right\} $,
where the 'actions' $\mathfrak{L}_{p}$ are defined as: $\mathfrak{L}_{p}\left(E,\Phi _{p},\mathfrak{J}_{p}\right)=\Phi _{p}^{\dagger }\left(\mathfrak{D}_{p}+\mathfrak{J}_{p}\right)\Phi _{p}$
in terms of the block-diagonal $4\times 4$ symmetric supermatrices
{[}7-10{]}:\[
\mathfrak{D}_{p}=\left(EI-H\right)\otimes L_{p}+i\left(\varepsilon I+\Gamma \right)\otimes \Lambda _{p}L_{p},\]
\[
L_{1}=\textrm{diag}\left\{ I_{2},I_{2}\right\} ,L_{2}=\textrm{diag}\left\{ I_{2},-I_{2}\right\} ,\]
\[
\Lambda _{1}=\textrm{diag}\left\{ I_{2},I_{2}\right\} ,\Lambda _{2}=\textrm{diag}\left\{ -I_{2},-I_{2}\right\} ,\]
\[
\mathfrak{J}_{1}=\textrm{diag}\left\{ 0,0,J^{\left(1\right)},J^{\left(2\right)}\right\} ,\mathfrak{J}_{2}=\textrm{diag}\left\{ 0,0,J^{\left(3\right)},J^{\left(4\right)}\right\} ,\]
Here $J^{\left(p\right)}$ are $N\times N$ symmetric source matrices,
and the integration measure is just a product of independent differentials
of commuting and anticommuting variables. The generating function
for the power moments $T=\mathfrak{D}_{ij}^{-1\, }\mathfrak{D}_{\, \, \, i\, j}^{\ast -1}$
and $T^{2}=\left(\mathfrak{D}_{ij}^{-1\, }\mathfrak{D}_{\, \, \, i\, j}^{\ast -1}\right)^{2}$
then can be shown to have the following representation:\begin{align}
Z\left(E,\mathfrak{J}\right)= & Z_{1}\left(E,\mathfrak{J}_{1}\right)Z_{2}\left(E,\mathfrak{J}_{2}\right)\notag \\
 & =\int \left[d\Phi \right]\exp \left\{ \frac{i}{2}\mathfrak{L}\, \left(E,\Phi ,\mathfrak{J}\right)\right\} ,\label{genfunc}
\end{align}
in terms of $8\times 8$ block-diagonal supermatrices $\mathfrak{D}=\textrm{diag}\left\{ \mathfrak{D}_{1},\mathfrak{D}_{2}\right\} ,\, L=\textrm{diag}\left\{ L_{1},L_{2}\right\} ,\, \Lambda =\textrm{diag}\left\{ \Lambda _{1},\Lambda _{2}\right\} ,\, \mathfrak{J}=\textrm{diag}\left\{ \mathfrak{J}_{1},\mathfrak{J}_{2}\right\} $
and $\Phi =\left\{ \Phi _{1},\Phi _{2}\right\} ,$ $\mathfrak{L}\left(E,\Phi ,\mathfrak{J}\right)=\mathfrak{L}_{1}\left(E,\Phi _{1},\mathfrak{J}_{1}\right)+\mathfrak{L}_{2}\left(E,\Phi _{2},\mathfrak{J}_{2}\right)=\Phi ^{\dagger }\left(\mathfrak{D}+\mathfrak{J}\right)\Phi $.

The Gaussian integral over the supervectors in Eq. (\ref{genfunc})
can be also written as a superdeterminant\[
Z\left(E,\mathfrak{J}\right)=\prod _{p=1,2}Z_{p}\left(E,\mathfrak{J}_{p}\right)=\prod _{p=1,2}\textrm{Sdet}\, \, ^{-1}\left(\mathfrak{D}_{p}+\mathfrak{J}_{p}\right).\]
Differentiating this expression with respect to elements of the symmetric
source matrix $\mathfrak{J}$ one finds (cf. \cite{VWZ,GuhrReview}):\begin{equation}
\frac{\partial ^{2}Z\left(E,\mathfrak{J}=0\right)}{\partial J_{ij}^{\left(1\right)}\partial J_{ij}^{\left(3\right)}}=T,\end{equation}
\begin{equation}
\frac{\partial ^{4}Z\left(E,\mathfrak{J}=0\right)}{\partial J_{ij}^{\left(1\right)}\partial J_{ij}^{\left(2\right)}\partial J_{ij}^{\left(3\right)}\partial J_{ij}^{\left(4\right)}}=T^{2},\end{equation}
relating both $T$ and $T^{2}$ to the Gaussian integrals over the
supervector components. Using the shorthand notation $\left\langle ...\right\rangle _{\Phi }=\int \left[d\Phi \right]\left(\ldots \right)\exp \left\{ i\mathfrak{L}\, \left(E,\Phi \right)/2\right\} $,
we can write\begin{equation}
T=\left\langle F_{1}\left[\Phi \right]\right\rangle _{\Phi },\end{equation}
\begin{equation}
T^{2}=\left\langle F_{2}\left[\Phi \right]\right\rangle _{\Phi },\end{equation}
where we introduced the following products of the commuting components
of the supervectors\[
F_{1}\left[\Phi \right]=S\left(1\right)_{1i}S\left(1\right)_{2i}S\left(1\right)_{1j}S\left(1\right)_{2j},\]
\begin{align*}
F_{2}\left[\Phi \right] & =S\left(1\right)_{1i}S\left(1\right)_{2i}S\left(2\right)_{1i}S\left(2\right)_{2i}\\
 & \times S\left(1\right)_{1j}S\left(1\right)_{2j}S\left(2\right)_{1j}S\left(2\right)_{2j}.
\end{align*}
Now, we proceed with GOE averaging of the above expressions for the
moments of the transmitted power. In what follows we use the overbar
to denote the averaging over $H$ with the weight $\exp \left\{ -\left(N/4v^{2}\right)TrH^{T}H\right\} $,
so that $\overline{H_{ij}H_{kl}}=\left(v^{2}/N\right)\left(\delta _{ik}\delta _{il}+\delta _{il}\delta _{jk}\right)$,
i.e. the ensemble averaging. It can be performed exactly with the
help of the identity:\[
\overline{\exp \left\{ \frac{i}{2}\Phi ^{\dagger }\left(H\otimes L\right)\Phi \right\} }=\exp \left\{ -\frac{v^{2}}{4N}\textrm{Str}A^{2}\right\} ,\]
where we introduced a new $8\times 8$ supermatrix: $A=L^{1/2}\sum _{i=1}^{N}\Phi _{i}\Phi _{i}^{\dagger }L^{1/2}.$
The elements of the supermatrix $A$ are labeled as follows:\[
A=\left(\begin{array}{cc}
 A_{mn}^{11} & A_{mn}^{12}\\
 A_{mn}^{21} & A_{mn}^{22}\end{array}
\right),\]
where $m,n=1,\ldots ,4.$. With the help of these notations we can
express $\overline{T}$ and $\overline{T^{2}}$ in a unified form
via the representations\begin{widetext}
\[
\overline{\left\langle F_{1,2}\left[\Phi \right]\right\rangle _{\Phi }}=\int \left[d\Phi \right]F_{1,2}\left[\Phi \right]\exp \left\{ \frac{i}{2}E\Phi ^{\dagger }L\Phi -\frac{1}{2}\Phi ^{\dagger }\left(\Gamma \otimes \Lambda \right)L\Phi -\frac{v^{2}}{4N}\textrm{Str}A^{2}-\frac{\varepsilon }{2}\textrm{Str}A\Lambda \right\} \]
\end{widetext}as both formulas differ only in the form of preexponent factors $F$.

\subsection{Performing $\Phi $-integration}

The next step of the supersymmetric calculation is the so-called Hubbard-Stratonovich
decoupling \cite{Efetovbook,GuhrReview}:\begin{align*}
\exp  & \left\{ -\frac{v^{2}}{4N}\textrm{Str}A^{2}-\frac{\varepsilon }{2}\textrm{Str}A\Lambda \right\} \\
= & \int \left[dR\right]\exp \left\{ -\frac{N}{4}\textrm{Str}R^{2}+i\frac{\varepsilon }{2v}N\textrm{Str}R\Lambda +i\frac{v}{2}\textrm{Str}RA\right\} ,
\end{align*}
\begin{align}
\overline{\left\langle F_{1,2}\left[\Phi \right]\right\rangle _{\Phi }}= & \int \left[dR\right]\exp \left\{ -\frac{N}{4}\textrm{Str}R^{2}+i\frac{\varepsilon }{2v}N\textrm{Str}R\Lambda \right\} \notag \\
\times \int \left[d\Phi \right] & F_{1,2}\left[\Phi \right]\exp \left\{ -\frac{i}{2}\Phi ^{\dagger }L^{1/2}f^{-1}L^{1/2}\Phi \right\} ,\label{F12bar}
\end{align}
where we defined $8N\times 8N$ supermatrix $f$:\begin{align*}
f & =\left[-EI\otimes I_{8}-vI\otimes R-i\left(\Gamma \otimes \Lambda \right)\right]^{-1}\\
 & =\left[\left(I_{N}\otimes I_{8}-i\Gamma \otimes \left(\Lambda \mathfrak{G}^{-1}\right)\right)\mathfrak{G}\right]^{-1},
\end{align*}
with $\mathfrak{G}=-EI_{8}-vR.$ In Eq. (\ref{F12bar}) we can integrate
out $\Phi -$variables, using Wick's theorem for supervectors, and
bring the remaining integral into a form suitable for a saddle point
approximation in the limit $N\rightarrow \infty $ \cite{Efetovbook,GuhrReview}.
Then for $i,j>M$ we obtain:\begin{align}
\int \left[d\Phi \right]F_{1,2}\left[\Phi \right] & \exp \left\{ -\frac{i}{2}\Phi ^{\dagger }f^{-1}\Phi \right\} \notag \\
 & =F_{1,2}\left[\mathfrak{G}^{-1}\right]\left(\textrm{Sdet}f\right)^{1/2}.\label{phiint}
\end{align}
Here we introduced the notations\begin{equation}
F_{1}\left[\mathfrak{G}^{-1}\right]=\frac{1}{4}\left\{ \left(\mathfrak{G}^{-1}\right)_{33}^{12}+\left(\mathfrak{G}^{-1}\right)_{33}^{21}\right\} ^{2},\label{F1G}\end{equation}
and\begin{align}
F_{2}\left[\mathfrak{G}^{-1}\right] & =\left\{ \left(\mathfrak{G}_{+}^{-1}\right)_{34}^{11}\left(\mathfrak{G}_{+}^{-1}\right)_{34}^{22}+\left(\mathfrak{G}_{+}^{-1}\right)_{33}^{12}\left(\mathfrak{G}_{+}^{-1}\right)_{44}^{12}\right.\notag \\
 & \left.+\left(\mathfrak{G}_{+}^{-1}\right)_{34}^{12}\left(\mathfrak{G}_{+}^{-1}\right)_{43}^{12}\right\} ,\label{F2G}
\end{align}
where $\mathfrak{G}_{+}^{-1}=\left\{ \mathfrak{G}^{-1}+\left(\mathfrak{G}^{-1}\right)^{T}\right\} /2$.
At this point we summarize the results for $\overline{T}$ and $\overline{T^{2}}$
separately:\begin{equation}
\overline{T}=\int \left[dR\right]F_{1}\left[\mathfrak{G}^{-1}\right]\exp \left\{ -N\mathcal{L}\left[R\right]+\delta \mathcal{L}\right\} ,\label{tbar2}\end{equation}
\begin{equation}
\overline{T^{2}}=\int \left[dR\right]F_{2}\left[\mathfrak{G}^{-1}\right]\exp \left\{ -N\mathcal{L}\left[R\right]+\delta \mathcal{L}\right\} ,\label{tsqbar2}\end{equation}
where the exponential is given by\begin{equation}
\mathcal{L}\left[R\right]=\frac{1}{4}\textrm{Str}R^{2}+\frac{1}{2}\textrm{Str}\ln \left(-EI_{8}-vR\right),\end{equation}
\begin{align}
\delta \mathcal{L}= & i\frac{\varepsilon }{2v}N\, \textrm{Str}R\Lambda \notag \\
 & -\frac{M}{2}\, \textrm{Str}\ln \left[I_{8}-i\gamma \Lambda \left(-EI_{8}-vR\right)^{-1}\right].
\end{align}
The remaining step is to carry out integration in Eqs. (\ref{tbar2},\ref{tsqbar2})
by the saddle point method in the limit of large $N$. The stationarity
condition for $\mathcal{L}\left[R\right]$ yields the saddle point
equation $R_{s}=v/\left(-EI_{8}-vR_{s}\right)$. Its solution is given
by a saddle-point manifold in a space of $8\times 8$ supermatrices
\cite{Efetovbook,VWZ}:\begin{equation}
R_{s}=-\frac{E}{2v}I_{8}+i\pi \nu v\mathfrak{T}^{-1}\Lambda \mathfrak{T}=-\frac{E}{2v}I_{8}-\pi v\nu Q.\label{saddlepoint}\end{equation}
Here $\nu $ denotes the mean eigenvalue density given for the GOE
by the Wigner semicircular law $\nu =\sqrt{4v^{2}-E^{2}}/\left(2\pi v^{2}\right)$.
After integrating out the massive Gaussian fluctuations around the
saddle point manifold in Eqs. (\ref{tbar2},\ref{tsqbar2}), the first
two moments of the transmitted power are expressed as integrals over
the supermatrices $Q=\mathfrak{T}^{-1}\Lambda \mathfrak{T}$ \cite{Efetovbook,VWZ}:\begin{align}
\frac{\overline{T}}{\left(\pi \nu \right)^{2}}= & \left\langle F_{1}\left[Q\right]\right\rangle _{Q}=\int \left[dQ\right]F_{1}\left[Q\right]\notag \\
\times  & \textrm{Sdet}^{-M/2}\left[I_{8}+i\frac{E}{2v^{2}}\gamma \Lambda +i\pi \nu \gamma Q\Lambda \right]\notag \\
\times  & \exp \left\{ -\frac{i}{2}\varepsilon \pi \nu N\textrm{Str}Q\Lambda \right\} ,\label{tbar3}
\end{align}
\begin{equation}
\frac{\overline{T^{2}}}{\left(\pi \nu \right)^{4}}=\left\langle F_{2}\left[Q\right]\right\rangle _{Q}.\label{tsqbar3}\end{equation}
This step completes derivation of the zero-dimensional nonlinear $\sigma $-model.

\subsection{Performing $Q$-integration}

To evaluate the superintegrals in Eqs. (\ref{tbar3},\ref{tsqbar3}),
we need to calculate $F_{1}\left[Q\right]$ and $F_{2}\left[Q\right]$
first. At this point we employ the Verbaarschot-Weidenmüller-Zirnbauer
(VWZ) parameterization \cite{VWZ} for the matrix $Q$. Both $F_{1}\left[Q\right]$
and $F_{2}\left[Q\right]$ are the functions of matrix elements of
$Q$, obtained by the formal substitution of $Q$ for $G^{-1}$ in
Eqs. (\ref{F1G},\ref{F2G}), as follows from Eq. (\ref{saddlepoint}).
Matrix elements of $Q$ are, in turn, the functions of eight commuting
and eight anticommuting variables. Although we are interested in the
highest order terms in anticommuting variables \cite{Gossaiaux1},
the calculation of $F_{1}\left[Q\right]$ and $F_{2}\left[Q\right]$
is too cumbersome to be done by hand. The calculation can be managed
most efficiently by employing the symbolic manipulation package epicGRASS
\cite{HartmannDavis}. The outputs of the epicGRASS (the highest order
terms in anticommuting variables) need to be further integrated over
all the anticommuting variables, and finally over all the commuting
variables except 'eigenvalues' \cite{VWZ}. After those steps we change
to the $\lambda $-variables of Ref. \cite{Efetovbook}, and arrive
at the representation for $T$ and $T^{2}$ in terms of a three-fold
integral. The details of this procedure are outlined in the Appendix
A. Here we only give the final expression:\begin{align}
 & \frac{\overline{T}}{\left(\pi \nu \right)^{2}}=\int _{1}^{\infty }d\lambda _{1}\int _{1}^{\infty }d\lambda _{2}\int _{-1}^{1}d\lambda F_{1}\left(\lambda ,\lambda _{1},\lambda _{2}\right)\notag \\
 & \times \exp \left\{ -\epsilon \left(\lambda _{1}\lambda _{2}-\lambda \right)\right\} \mu \left(\lambda ,\lambda _{1},\lambda _{2}\right)\Pi \left(\lambda ,\lambda _{1},\lambda _{2}\right),\label{tbar4}
\end{align}
\begin{align}
 & \frac{\overline{T^{2}}}{\left(\pi \nu \right)^{4}}=\int _{1}^{\infty }d\lambda _{1}\int _{1}^{\infty }d\lambda _{2}\int _{-1}^{1}d\lambda F_{2}\left(\lambda ,\lambda _{1},\lambda _{2}\right)\notag \\
 & \times \exp \left\{ -\epsilon \left(\lambda _{1}\lambda _{2}-\lambda \right)\right\} \Pi \left(\lambda ,\lambda _{1},\lambda _{2}\right)\mu \left(\lambda ,\lambda _{1},\lambda _{2}\right),\label{tsqbar4}
\end{align}
where $\epsilon =2\pi \nu N\varepsilon $, and\[
\mu \left(\lambda ,\lambda _{1},\lambda _{2}\right)=\frac{1-\lambda ^{2}}{\left(\lambda _{1}^{2}+\lambda _{2}^{2}+\lambda ^{2}-2\lambda \lambda _{1}\lambda _{2}-1\right)^{2}},\]
\[
F_{1}\left(\lambda ,\lambda _{1},\lambda _{2}\right)=1-\lambda ^{2}+\left(\lambda _{1}^{2}-1\right)\lambda _{2}^{2}+\left(\lambda _{2}^{2}-1\right)\lambda _{1}^{2},\]
\[
F_{2}\left(\lambda ,\lambda _{1},\lambda _{2}\right)=2\left(1-\lambda _{1}^{2}-\lambda _{2}^{2}-2\lambda \lambda _{1}\lambda _{2}+3\lambda _{1}^{2}\lambda _{2}^{2}\right)^{2}.\]
The remaining factor $\Pi \left(\lambda ,\lambda _{1},\lambda _{2}\right)$
contains all the information about the dissipation channels and comes
from a calculation of the relevant superdeterminant (cf. \cite{FyodorovSommers}
for the GUE case)\begin{widetext}\begin{align}
S\det \, ^{-M/2}\left[I_{8}+i\frac{E}{2v^{2}}\gamma \Lambda +i\pi \nu \gamma Q\Lambda \right]\qquad \qquad \qquad \qquad \qquad \qquad \qquad \quad  & \notag \\
=\left(\frac{v^{2}+\gamma ^{2}+2\pi \nu \gamma \lambda }{\sqrt{\left(v^{2}+\gamma ^{2}\right)^{2}+4\pi \nu \gamma v^{2}\left(v^{2}+\gamma ^{2}\right)\lambda _{1}\lambda _{2}+\left(2\pi \nu \gamma v^{2}\right)^{2}\left(\lambda _{1}^{2}+\lambda _{2}^{2}-1\right)}}\right)^{M} & \notag \\
=\frac{\left(g+\lambda \right)^{M}}{\left(\sqrt{g^{2}+2g\lambda _{1}\lambda _{2}+\lambda _{1}^{2}+\lambda _{2}^{2}-1}\right)^{M}}=\Pi \left(\lambda ,\lambda _{1},\lambda _{2}\right),\qquad \qquad \qquad \qquad \quad  &
\end{align}
\end{widetext}where $g=\left(v^{2}+\gamma ^{2}\right)/\left(2\pi \nu \gamma v^{2}\right)$,
and we have also used:\begin{equation}
StrQ\Lambda =-4i\left(\lambda _{1}\lambda _{2}-\lambda \right).\end{equation}
 To generalize Eqs. (\ref{tbar4},\ref{tsqbar4}) to the case of non-equipotent
dampers we just need to replace $\Pi \left(\lambda ,\lambda _{1},\lambda _{2}\right)$
with\[
\Pi \left(g_{i},\lambda ,\lambda _{1},\lambda _{2}\right)=\prod _{i}\frac{\left(g_{i}+\lambda \right)}{\left(\sqrt{g_{i}^{2}+2g_{i}\lambda _{1}\lambda _{2}+\lambda _{1}^{2}+\lambda _{2}^{2}-1}\right)},\]
see, for example \cite{VWZ,channels}. It can be verified that Eq.
(\ref{tbar4}) yields the same result for $\overline{T}$ as follows
from adopting the final formula of Ref. \cite{VWZ}.

\subsection{Special case of uniform damping. Comparison with naive calculation.}

Next, we compare results of the present (supersymmetric) calculation
with the results of Ref. \cite{LobkisJSV} for the case of uniform
damping $M=0,\epsilon \neq 0$. In that special case the naive calculation
of Ref. \cite{LobkisJSV} should be exact. In order to obtain $\overline{T}$
we need to evaluate the integral\begin{align}
I\left(x\right) & =\int _{1}^{\infty }d\lambda _{1}\int _{1}^{\infty }d\lambda _{2}\int _{-1}^{1}d\lambda \exp \left\{ ix\left(\lambda _{1}\lambda _{2}-\lambda \right)\right\} \notag \\
 & \qquad \qquad \quad \times F_{1}\left(\lambda ,\lambda _{1},\lambda _{2}\right)\mu \left(\lambda ,\lambda _{1},\lambda _{2}\right),
\end{align}
where we have denoted $x=i\epsilon $. The Fourier transformation
with respect to the $x-$variable,\[
\widetilde{I}\left(t\right)=\int _{-\infty }^{\infty }I\left(x\right)\exp \left\{ -ixt\right\} dx,\]
has a meaning of averaged response power in the time domain for a
system without dissipation. It can be written in a more convenient
form:\begin{align}
\widetilde{I}\left(t\right) & =2\pi \int _{1}^{\infty }d\lambda _{1}\int _{1}^{\infty }d\lambda _{2}\int _{-1}^{1}d\lambda \delta \left(\lambda -\lambda _{1}\lambda _{2}+t\right)\notag \\
 & \qquad \qquad \quad \times F_{1}\left(\lambda ,\lambda _{1},\lambda _{2}\right)\mu \left(\lambda ,\lambda _{1},\lambda _{2}\right)
\end{align}
After performing $\lambda -$integration, we make the change of variables:
$u=\lambda _{1}\lambda _{2},z=\lambda _{1}^{2}$ suggested in Ref.
\cite{Efetovbook}, and after a lengthy but straightforward procedure
arrive at a very simple expression:\[
\widetilde{I}\left(t\right)=4\pi \theta \left(t\right).\]
which can be immediately Fourier-inverted, yielding\[
I\left(x\right)=\frac{-2i}{x}\]
 This is equivalent to the first moment of the transmitted power given
by \begin{equation}
\frac{\overline{T}}{\left(\pi \nu \right)^{2}}=\frac{2}{\epsilon },\label{Tbaruniform}\end{equation}
 and indeed coincides with the value predicted by the naive calculation
of Ref. \cite{LobkisJSV}. 

The same steps can be repeated when calculating the second moment
$\overline{T^{2}}$. One starts with Fourier-transforming the right-hand
side of Eq. (\ref{tsqbar4}), then changes to the variables $u$ and
$z$, carries out the remaining double integration explicitly and
finally applies the Fourier-inversion. Intermediate calculations are
too long to be reproduced in the paper, but the final result reads:\begin{align}
\frac{\overline{T^{2}}}{\left(\pi \nu \right)^{4}} & =\frac{1}{\epsilon ^{4}}\left(5+28\epsilon +7\epsilon ^{2}\right)-\frac{e^{-2\epsilon }}{\epsilon ^{4}}\left(5+2\epsilon +\epsilon ^{2}\right)\notag \\
 & \qquad +\frac{e^{-\epsilon }}{\epsilon ^{4}}E_{1}\left(\epsilon \right)\left(10+10\epsilon +3\epsilon ^{2}+\epsilon ^{3}\right)\notag \\
 & \qquad +\frac{e^{\epsilon }}{\epsilon ^{4}}E_{1}\left(\epsilon \right)\left(-10+10\epsilon -3\epsilon ^{2}+\epsilon ^{3}\right),\label{Tsqbaruniform}
\end{align}
where\[
E_{1}\left(z\right)=\int _{z}^{\infty }\frac{e^{-s}}{s}ds.\]
This matches perfectly with the corresponding result of Ref. \cite{LobkisJSV}.

\section{Numerical results for the moments of the transmitted power.}

The predictions Eqs. (\ref{tbar4},\ref{tsqbar4}) of the supersymmetric
calculations can be compared with Monte Carlo evaluations of the first
two moments of $T$. Towards this goal we numerically generated an
ensemble of $N\times N$ real symmetric matrices $H$ typically choosing
$1500$ ensemble realizations and taking $N=1000$. The procedure
is almost identical to that described in Ref. \cite{GUEpaper}. The
entries in $H$ are constructed using a Gaussian random number generator
such that $\overline{H_{ij}H_{kl}}=\left(1/N\right)\left(\delta _{ik}\delta _{jl}+\delta _{il}\delta _{jk}\right)$.
To simulate the case of the uniform damping we use $\Gamma =\varepsilon I$.
To simulate the case of a finite number of decay channels we take
the diagonal $\Gamma =\textrm{diag}\{\gamma ,\gamma ,...,\gamma ,0,...,0\}$
with $M<N$ identical positive entries. Then, for every ensemble realization
we generate the off-diagonal elements of the resolvent matrix according
to $G_{ij}(E)=\left[EI+i\Gamma -H\right]^{-1}$, modeling in this
way the response at a site $i$ due to excitation at the site $j$,
with $E$ standing for the spectral parameter, and both $i$ and $j$
chosen to be large than $M$ to avoid direct coupling to the damping
channels.%
\begin{figure}[htbp]
\begin{center}\includegraphics{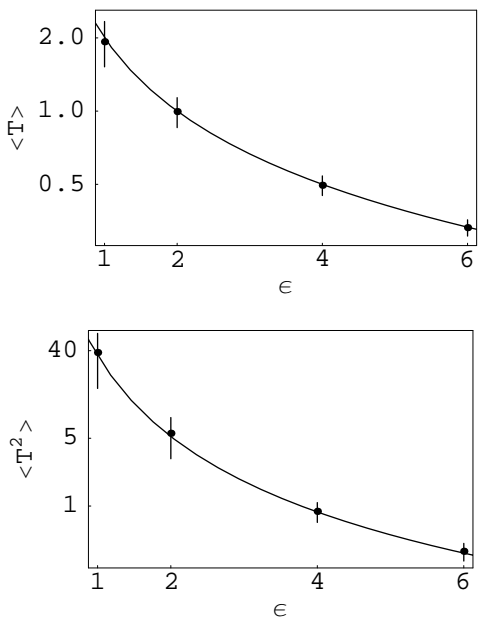}\end{center}

\caption{FIG. 1. $\left\langle T\right\rangle $ and $\left\langle T^{2}\right\rangle $
are plotted on log-scale as the functions of parameter $\epsilon $
for the case of uniform damping. The solid lines represent theoretical
predictions (Eqs. (\ref{tbar4}) and (\ref{tsqbar4})). For each numerically
obtained $\left\langle T\right\rangle $ and $\left\langle T^{2}\right\rangle $(represented
by dots) $1500$ samples of $\left|G_{ij}\left(E\right)\right|^{2},\; i\neq j$
were computed. Five sigma error bars were computed based on the observed
variances of $T$ and $T^{2}$. }
\end{figure}

Let us first consider the case of the uniform damping: $\Gamma =\varepsilon I$.
For a fixed matrix size $N$ and fixed value of the spectral variable
$E$ we explore a range of $\varepsilon $. For $E=0$ the modal density
$\partial N/\partial E$ is given by $\nu =1/\pi $. Mean level width
$\overline{\gamma }=2\pi \nu \left\langle \Im E_{r}\right\rangle $
in this case is identical to $\epsilon =2\pi \nu N\varepsilon $.
In Fig. 1 we compare both moments of power $T$ as given by Eqs. (\ref{Tbaruniform},\ref{Tsqbaruniform})
with the results of Monte Carlo simulations for several values of
$\epsilon $. It is evident that numerical results correspond well
with the theoretical curves. 

To repeat the same procedure for finite number of local dampers $M$
we evaluated the three-dimensional integrals in Eqs. (\ref{tbar4},\ref{tsqbar4})
numerically for a broad range of the scaled mean level width $\overline{\gamma }$
\cite{GUEpaper}. The difficulties of the numerical integration arising
due to the singularity of $\mu \left(\lambda ,\lambda _{1},\lambda _{2}\right)$,
are easy to overcome by employing the change of variables suggested
in Ref. \cite{VerbaarschotAnnPhys}. The results are presented in
Fig. 2 and also show a good agreement with the theory.%
\begin{figure*}
\begin{center}\includegraphics{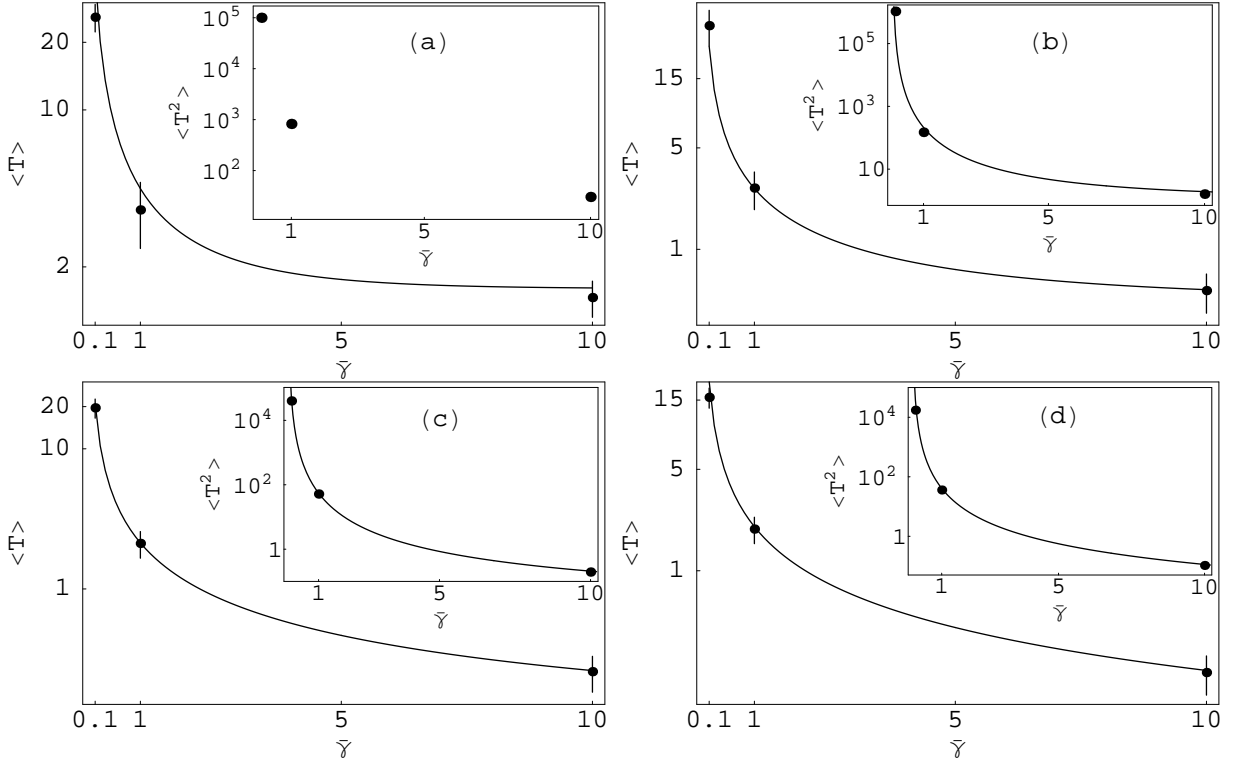}\end{center}

\caption{FIG. 2. Mean power (main figure) and mean square power (inset) for:
(a) $M=4$; (b) $M=10$; (c) $M=40$; (d) $M=400$. Solid lines represent
theoretical predictions (Eqs. (\ref{tbar4}) and (\ref{tsqbar4})).
For each numerically obtained $\left\langle T\right\rangle $ and
$\left\langle T^{2}\right\rangle $ (dots) $1500$ samples of $\left|G_{ij}\left(E\right)\right|^{2},$
were computed. We imposed the restrictions: $i\neq j$, and $i>M,j>M$
for the non-uniform damping case, to avoid 'recording' the response
from damped sites or from the 'source' site $j$, and to correspond
to the assumptions in the theoretical analysis. For the numerically
obtained mean power, twenty, ten and five sigma error bars were computed
for the scaled mean level width $\overline{\gamma }=0.1,1.0,10.0$
respectively. They were based on the observed variances of $T$ and
$T^{2}$. Error bars for the power variance are not shown, because
they are smaller than the size of the dots. For $M=4$ the theoretical
prediction for the variance does not exist.}
\end{figure*}

\section{Relative variance}

Dissipation within the framework of the present approach is parametrized
in terms of quantities $g$, $M$ and $\epsilon $. At the same time
those quantities are not readily accessible experimentally, and in
any case were not measured in the work reported in Ref. \cite{LobkisJSV}.
For this reason any direct comparison with those measurements is not
feasible. Nevertheless, by choosing plausible values for the relevant
parameters we can investigate the sign and magnitude of the discrepancies
arising between the predictions for the relative variance of the transmitted
power calculated in the two theories under discussion. As a result
of such comparison we found that the discrepancy between the naive
analysis and the present (supersymmetric) calculation is similar to
one reported previously in \cite{LobkisJSV} between the naive theory
and actual measurements. 

The comparison is carried out by first considering the mean spectral
energy density (mean square response) in the time domain: $E(t)=\left|G_{ij}\left(t\right)\right|^{2}$,
where $G_{ij}\left(t\right)$ is the (band-limited) time-domain Green's
function. Similar quantities were studied previously in the context
of the delay time distributions in chaotic cavities \cite{DittesReview}.
Their statistics can be obtained from the inverse Fourier transform
of the two-point correlation function $T\left(\Omega \right)=\overline{G_{ij}\left(E\right)G_{ij}^{*}\left(E+\Omega \right)}$
\cite{VWZ,VerbaarschotAnnPhys} with respect to $\Omega $. The expression
for $T\left(\Omega \right)$ can be obtained by replacing $2i\varepsilon $
with $\Omega $ in the derivation of Eq. (\ref{tbar4}), (see also
\cite{VerbaarschotAnnPhys}). Thus,\begin{eqnarray}
 & E(\tau )\sim \int _{1}^{\infty }\int _{1}^{\infty }d\lambda _{1}d\lambda _{2}\Pi \left(\tau ,\lambda _{1},\lambda _{2}\right)f\left(\tau ,\lambda _{1},\lambda _{2}\right) & \notag \\
 & \times \frac{\theta \left(\lambda _{1}\lambda _{2}-\tau +1\right)\theta \left(\tau -\lambda _{1}\lambda _{2}+1\right)\left(1-\left(\tau +\lambda _{1}\lambda _{2}\right)^{2}\right)}{\left(\lambda _{1}^{2}+\lambda _{2}^{2}+\left(\tau +\lambda _{1}\lambda _{2}\right)^{2}-2\lambda _{1}\lambda _{2}\left(\tau +\lambda _{1}\lambda _{2}\right)-1\right)^{2}}, & \label{eoftau}
\end{eqnarray}
where $\tau =t/\left(2\pi \nu N\right)$ is a dimensionless time,
$f\left(\tau ,\lambda _{1},\lambda _{2}\right)=\left(\lambda _{1}^{2}-1\right)\lambda _{2}^{2}+\left(\lambda _{2}^{2}-1\right)\lambda _{1}^{2}+1-\left(\tau +\lambda _{1}\lambda _{2}\right)^{2}$,
and $\Pi \left(\tau ,\lambda _{1},\lambda _{2}\right)=\left(g+2\tau +\lambda _{1}\lambda _{2}\right)^{M}\left(g^{2}+2g\lambda _{1}\lambda _{2}+\lambda _{1}^{2}+\lambda _{2}^{2}-1\right)^{-M/2}$.
The generalization to the case of non-equipotent channels is straightforward:
$\Pi \left(\tau ,\lambda _{1},\lambda _{2}\right)=\prod _{i}^{M}\left(g_{i}+2\tau +\lambda _{1}\lambda _{2}\right)\left(g_{i}^{2}+2g_{i}\lambda _{1}\lambda _{2}+\lambda _{1}^{2}+\lambda _{2}^{2}-1\right)^{-1/2}$.
The naive method yields a simpler expression for spectral energy density
\cite{LobkisJSV,DittesReview,Burkhardt}:\begin{equation}
E(\tau )_{naive}=E_{0}\left(1+\frac{2\sigma \tau }{M}\right)^{-M/2},\label{et2}\end{equation}
where the initial logarithmic decay rate $\sigma $ is proportional
to the mean resonance width, given by a Porter-Thomas distribution. 

In Ref. \cite{LobkisJSV} $E(\tau )$ was measured experimentally,
and fitted into the naive result (\ref{et2}) to extract values for
$M$ and $\sigma $. The two parameters were further used to predict
the relative variance of $T$ ($relative\, variance=\left\langle T^{2}\right\rangle /\left\langle T\right\rangle ^{2}-1$).
Having the exact result (Eq. (\ref{eoftau})) we can now attempt to
explain the $20-30\%$ over-prediction of relative variance reported
in Ref. \cite{LobkisJSV}. Clearly, by specifying certain values for
$M$, $g$ and $\epsilon $ the wave scattering in an ergodic sample
can be completely described, since both spectral energy density and
relative variance are fixed uniquely. Further assuming that $E(\tau )$
as given by Eq. (\ref{eoftau}) is the {}``measured'' energy density
of our system, we can repeat the procedure of Ref. \cite{LobkisJSV}.
Namely, we fit it to $E(\tau )_{naive}$ in order to calculate relative
variance according to the two-parameter naive formula used in \cite{LobkisJSV}
for comparison with actual measurements. Such a fit allows to extract
values for $M_{naive}$, $\sigma _{naive}$ and $E_{0}$ that may,
or may not, correspond to the exact values. The true value of relative
variance as determined from Eqs. (\ref{tbar4},\ref{tsqbar4}) may
then be compared to the corresponding naive prediction. 

By a numerical three-parameter fit over the same dynamic range (of
a factor of $e^{10}$) as in \cite{LobkisJSV}, we obtained values
for $E_{0},$ $\sigma _{naive}$ and $M_{naive}$. In spite of the
naivete of the model the fits were generally quite good, as observed
in \cite{LobkisJSV}, and we can substitute the obtained values into
the formula for the relative variance from Ref. \cite{LobkisJSV}\begin{align}
\frac{\left\langle T^{2}\right\rangle }{\left\langle T\right\rangle ^{2}}-1 & =1+\frac{9}{\sigma }\frac{M\left(M-2\right)}{\left(M-4\right)\left(M-6\right)}\notag \\
 & -4\left\{ i_{1}+\sigma ^{2}\frac{\left(M-2\right)^{2}}{M^{2}}i_{2}\right\} ,\label{relvar}
\end{align}
where\[
i_{1}=\frac{M}{2\sigma }\exp \left\{ \frac{M}{\sigma }\right\} E_{M-2}\left(\frac{M}{\sigma }\right),\]
\begin{align*}
i_{2} & =\frac{M^{3}}{8\sigma ^{3}}\exp \left\{ \frac{M}{\sigma }\right\} \left\{ E_{M}\left(\frac{M}{\sigma }\right)\right.\\
 & \left.-2E_{M-1}\left(\frac{M}{\sigma }\right)+E_{M-2}\left(\frac{M}{\sigma }\right)\right\} ,
\end{align*}
\[
E_{k}\left(z\right)=\int _{1}^{\infty }\frac{e^{-zs}}{s^{k}}ds.\]
\begin{table}[htbp]
\begin{tabular}{|c|c|c|c|c|}
\hline 
$M$&
$g$&
$\sigma $&
naive&
exact\\
\hline
\hline 
10&
20.017&
0.497&
59.881&
59.492\\
\hline 
20&
20.017&
0.989&
14.419&
14.397\\
\hline 
20&
10.033&
1.930&
7.582&
7.574\\
\hline 
14&
2.918&
4.066&
4.990&
4.776\\
\hline
\end{tabular}

\caption{Relative variance in absence of overall damping.}
\end{table}
\begin{table}[htbp]
\begin{tabular}{|c|c|c|c|c|c|}
\hline 
$M$&
$g$&
$\epsilon $&
$\sigma $&
naive&
exact\\
\hline
\hline 
1&
1.0&
1.0&
1.302&
7.711&
6.801\\
\hline 
1&
2.0&
1.0&
1.205&
7.908&
6.809\\
\hline 
1&
5.0&
1.0&
1.126&
8.195&
6.784\\
\hline 
4&
10.0&
1.0&
1.330&
7.194&
6.134\\
\hline 
6&
9.0&
0.5&
1.008&
10.557&
8.611\\
\hline
\end{tabular}

\caption{Relative variance in presence of overall damping. }
\end{table}
The results for several values of parameters are summarized in Tab.
1 and Tab. 2. It appears that in the absence of overall damping (Tab.
1) the actual value of relative variance is very close to its naive
estimate. However, when we consider the case of a small number of
strong dampers in a system with a uniform background $\epsilon \ne 0$
(Tab. 2), the difference becomes similar to the discrepancy found
in \cite{LobkisJSV}. A more definitive comparison of Eqs. (\ref{tbar4},\ref{tsqbar4})
with measurements awaits an experiment in which the values of $\epsilon $
and the $g_{i}$ can be ascertained independently.

\section{Conclusions}

In the present paper the special cases of two and four point correlation
functions of the transmitted power spectrum have been calculated both
analytically and numerically for ergodic dissipative structures. In
the context of the wave scattering the former corresponds to the mean,
and the latter to the mean square of the transmitted wave power $T$. 

The ergodicity assumption is implicit by virtue of our replacement
of the actual differential operator describing wave motion by a large
random symmetric matrix. Dissipation is taken to act both locally
in space ({}``localized dampers'' or dissipative channels) and uniformly
within the sample. 

In accord with earlier results \cite{GUEpaper}, the presence of nonuniform,
or localized, sources of dissipation requires the use of an elaborate
nonperturbative technique --- the so-called zero dimensional supersymmetric
non-linear $\sigma $-model --- to obtain the moments of the transmitted
power. It is found that the naive approach fails to correctly describe
mean square power; the failure is ascribable at least in part to the
assumption of real Gaussian eigenmodes inherent in that approach.
The supersymmetry technique allows one to bypass the difficulty of
identifying eigenmode statistics, and to arrive at expressions which
are in agreement with Monte Carlo simulations, and appear to be in
better agreement with experimentally measured values of variance. 

\begin{acknowledgments}
This work was supported by grants from the National Science Foundation
{[}CMS 99-88645 and CMS 0201346{]}, by computational resources from
the National Center of Supercomputing Applications and by Vice-Chancellor
grant from Brunel University. IR would like to thank International
Programs in Engineering in University of Illinois at Urbana-Champaign
for financial support, the Department of Mathematical Sciences at
Brunel University for financial support and hospitality during his
visit. 
\end{acknowledgments}

\appendix

\section{Evaluation of the superintegral}

In this Appendix we elaborate on steps leading to the main results
of Section II, expressed by Eqs. (\ref{tbar4},\ref{tsqbar4}). We
start by evaluating $F_{1}\left[Q\right]$ and $F_{2}\left[Q\right]$
with epicGRASS. The program extracted terms of lowest and highest
order in anticommuting variables, which are, generally, the only terms
needed. The lowest order term was found to be unimportant since the
resulting integrands are not singular at the boundary \cite{Efetovbook,Gossaiaux1}.
Then, we simplified the output of epicGRASS with $Mathematica$ and
reduced the superintegral to a multiple integral over commuting and
anticommuting variables \cite{VWZ}. 

The elements of matrix $Q$ are introduced into the epicGRASS in terms
of the parametrization of Ref. \cite{VWZ}. Eight commuting variables
are: the {}``eigenvalues'' $\mu _{1},\mu _{2},\mu $, the parameters
of $SU\left(2\right)$ group $m,r,s$, and two {}``angles'' $\varphi _{1}$
and $\varphi _{2}$. The integration region in Eqs. (\ref{tbar3},\ref{tsqbar3})
corresponds to $-\infty <\mu _{1},\mu _{2},m,r,s<\infty $, $0<\mu <1$,
$0<\varphi _{1},\varphi _{2}<2\pi $ \cite{Section8ofVWZ}. Then,
after epicGRASS extracts the highest order term in anticommuting variables,
we have, for example, for $F_{1}\left[Q\right]$:\begin{align*}
F_{1}\left[Q\right] & =-32z^{2}-32z_{1}z_{2}\cos \varphi _{1}\cos \varphi _{2}\sin \varphi _{1}\sin \varphi _{2}\\
 & -z_{1}^{2}\left(36\cos \varphi _{1}^{2}\cos \varphi _{2}^{2}+12\cos \varphi _{1}^{2}\sin \varphi _{2}^{2}\right.\\
 & \left.+12\cos \varphi _{2}^{2}\sin \varphi _{1}^{2}+4\sin \varphi _{1}^{2}\sin \varphi _{2}^{2}\right)\\
 & -z_{2}^{2}\left(36\sin \varphi _{1}^{2}\sin \varphi _{2}^{2}+12\cos \varphi _{1}^{2}\sin \varphi _{2}^{2}\right.\\
 & \left.+12\cos \varphi _{2}^{2}\sin \varphi _{1}^{2}+4\cos \varphi _{1}^{2}\cos \varphi _{2}^{2}\right),
\end{align*}
where $z_{1,2}=\mu _{1,2}\sqrt{1+\mu _{1,2}^{2}}$, and $z=i\mu \sqrt{1-\mu ^{2}}$. 

Eight anticommuting variables are readily integrated out according
to the convention $\int d\chi \chi =1/\left(2\pi \right)^{1/2}$.
Note, that this convention is different from the one we took in the
beginning of Section II in the derivation of generating function.
However, this discrepancy has no influence on the remaining process,
as long as we use the integration measure of Ref. \cite{VWZ}. Finally,
integrating over the {}``angles'' as well as over the parameters
of $SU\left(2\right)$ we arrive at:\begin{equation}
\widetilde{F}_{1}\left[Q\right]=-16\left(z_{1}^{2}+z_{2}^{2}-2z^{2}\right),\label{F1tilde}\end{equation}
where we indicated the integration (which does not affect other factors
in the integrand in Eq. (\ref{tbar3})) by tilde. 

Upon the substitution of {}``eigenvalues'' into Eq. (\ref{F1tilde})
we can compare our expression for $\overline{T}$ with the final formula
of Ref. \cite{VWZ}. We switch to the combinations\[
\lambda _{1,2}^{V}=\mu _{1,2.}^{2},\, \lambda ^{V}=\mu ^{2},\]
which are the final variables appearing in the resulting expression
of Ref. \cite{VWZ}. Two results match perfectly, and we can proceed
with the corresponding calculation of the second moment of the transmitted
power. Before doing that we again change variables, this time ---
to the {}``eigenvalues'' of Efetov's parameterization, according
to\[
\lambda _{1,2}^{V}=\lambda _{1}\lambda _{2}\pm \sqrt{\left(\lambda _{1}^{2}-1\right)\left(\lambda _{2}^{2}-1\right)}.\]
The domain of the integration has to be modified as well: $1<\lambda _{1},\lambda _{2}<\infty ,-1<\lambda <1$.
The Efetov's {}``eigenvalues'' are somewhat more convenient for
the calculations done at the end of Section II, where we compared
the exact and naive results for the first two moments in uniform damping
case. 

The analogous procedure for $\overline{T^{2}}$ yields\begin{align}
\widetilde{F}_{2}\left[Q\right] & =4\left(4x^{2}-4xx_{1}-4xx_{2}+x_{1}^{2}+x_{2}^{2}\right.\notag \\
 & \left.+2x_{1}x_{2}+8z_{1}^{2}+8z_{2}^{2}-16z^{2}\right)^{2}\label{F2tilde}
\end{align}
where $x_{1,2}=1+2\mu _{1,2}^{2},x=1-2\mu ^{2}$, and after passing
to Efetov's variables in Eqs. (\ref{F1tilde},\ref{F2tilde}) we obtain
the final results of Section II --- Eqs. (\ref{tbar4},\ref{tsqbar4}).

\begin{thebibliography}{10}
\bibitem{Porter}C. E. Porter, \emph{Statistical Theories of Spectra: Fluctuations}
(Academic Press, New York, 1965); T. A. Brody \emph{et al}., Rev.
Mod. Phys. \textbf{53}(3), 385 (1981); P. A. Mello, in \textit{Mesoscopic
Quantum Physics}, \textit{\emph{Proceedings of Les Houches Summer
School}}, 1994, ed. by E. Akkermans \emph{et al}\textit{.,} \textit{\emph{(}}Elsevier,
Amsterdam, 1995), Session LXI, 435. 
\bibitem{BGS}O. Bohigas, M. J. Giannoni, and C. Schmit, Phys. Rev. Lett. \textbf{52},
1 (1984); H.-J. Stöckmann. \textit{Quantum Chaos: An Introduction}
\textit{\emph{(}}Cambridge University Press, Cambridge, 1999); O.
Bohigas, Random matrix theories and chaotic dynamics, in \textit{Mesoscopic
Quantum Physics}, \textit{\emph{Proceedings of Les Houches Summer
School}}, 1989, ed. by M.-J. Giannoni, A. Voros, and J. Zinn-Justin,
(Elsevier, Amsterdam, 1991), Session LII, 87. 
\bibitem{Kogan1}E. Kogan \emph{et al.} Phys. Rev. E \textbf{61}, R17 (2000); C. W.
J. Beenakker and P.W.Brouwer, Physica E \textbf{9}, 463 (2001); S.A.
Ramakrishna and N. Kumar, Phys. Rev. B \textbf{61}, 3163 (2000). 
\bibitem{GUEpaper}I. Rozhkov, Y. V. Fyodorov, and R. L. Weaver, Phys. Rev. E \textbf{68},
016204 (2003). 
\bibitem{MendezSnchez1}R.A. Mendez-Sanchez \emph{et al.} e-preprint cond-mat/0305090. 
\bibitem{SavinSommers1}D.V.Savin and H.-J.Sommers, Phys.Rev. E \textbf{68}, 036211 (2003) 
\bibitem{FyodorovJETP}Y.V. Fyodorov, JETP Letters \textbf{78}, 250 (2003) 
\bibitem{Lyon}R. H. Lyon, J. Acoust. Soc. Am. \textbf{45}, 545 (1969); J. L. Davy,
J. Sound Vib. \textbf{77}, 455 (1981); \textbf{107}, 361 (1986); \textbf{115},
145 (1987); M. R. Schroeder, J. Audio Eng. Soc. \textbf{10}, 219 (1962);
\textbf{37}, 409 (1965); \textbf{35}, 307 (1987); Proceedings of 5th
International Congress of Acoustics, G31 (1965). 
\bibitem{LobkisJSV}O.I. Lobkis, R.L. Weaver, I. Rozhkov, J. Sound Vib. \textbf{237}(2),
281 (2000). 
\bibitem{Efetovbook}K. Efetov, \textit{Supersymmetry in disorder and chaos} (Cambridge
University Press, 1997). 
\bibitem{VWZ}J. M. Verbaarschot, H.A. Weidenmüller, and M.R. Zirnbauer, Phys. Rep.
\textbf{129}, 367 (1985). 
\bibitem{GuhrReview}T. Guhr, A. Müller-Groeling, and H. A. Weidenmüller, Phys. Rep. \textbf{299},
189 (1998). 
\bibitem{FyodorovSommers}Yan V. Fyodorov and H.-J. Sommers, J. Math. Phys. \textbf{38} (4),
1918 (1997). 
\bibitem{AlhassidReview}Y. Alhassid, Rev. Mod. Phys. \textbf{72}, 895 (2000). 
\bibitem{DittesReview}F.-M. Dittes, Phys. Rep. \textbf{339}, 215 (2000). 
\bibitem{Burkhardt}J. Burkhardt, R. L.Weaver, J. Sound Vib. 196, 147 (1996); J. Burkhardt,
Ultrasonics \textbf{36}, 471 (1986); O.I. Lobkis, R.L. Weaver, and
I. Rozhkov, J. Sound Vib. \textbf{237}, 281 (2000); M. R. Schroeder,
5th Int. Congress of Acoustics, G31 (1965); O.I. Lobkis, I. Rozhkov,
and R.L. Weaver, e-print cond-mat//0305499 (Phys. Rev. Lett. to appear). 
\bibitem{VerbaarschotAnnPhys}J. J. M. Verbaarschot, Ann. Phys. \textbf{168}, 368 (1986). 
\bibitem{HartmannDavis}U. Hartmann and E. D. Davis, Comput. Phys. Commun. \textbf{54}, 353
(1989). 
\bibitem{Gossaiaux1}P.-B. Gossiaux, Z. Pluhar, and H.A. Weidenmüller, Ann. Phys. \textbf{268},
273 (1998). 
\bibitem{channels}Y. V. Fyodorov, D.Savin, and H.-J. Sommers, Phys. Rev. E \textbf{55},
R4857 (1977) ; Y. V. Fyodorov and Y. Alhassid, Phys. Rev. A \textbf{58},
R3375 (1998) 
\bibitem{SavinSokolovetal}D. V. Savin and V. V. Sokolov, Phys. Rev. E \textbf{56}, R4911 (1997);
H.-J. Sommers, D. V. Savin, and V. V. Sokolov, Phys. Rev. Lett. \textbf{87},
094101 (2001); P. W. Brouwer, K. M. Frahm, and C. W. J. Beenakker,
Phys. Rev. Lett. \textbf{78}, 4737 (1997); Waves Random Media \textbf{9},
91 (1999); M.G.A. Crawford and P.W. Brouwer, Phys. Rev E \textbf{65},
026221 (2002). 
\bibitem{Section8ofVWZ}Section 8 of Ref. \cite{VWZ} contains all the necessary information
on the VWZ-parameterization. 
\end{thebibliography}
\end{document}